\documentclass[vecphys]{svmult}

\usepackage{makeidx}         
\usepackage{graphicx}        
\usepackage{multicol}        
\usepackage[bottom]{footmisc}

\makeindex             

\begin{document}
 
\title*{Spectral Classification; Old and Contemporary}
\titlerunning{Spectral Classification} 
\author{Sunetra Giridhar}         
\institute{Indian Institute of Astrophysics, 
Bangalore 560034, India } 
%
%
\maketitle
\begin{abstract}
 Beginning with a historical account of the spectral classification,
 its refinement through additional criteria  is presented.
 The line strengths and ratios used in two dimensional classifications
 of each spectral class are described.
 A parallel classification scheme for metal-poor stars and
 the standards used for classification are presented.
 The extension of spectral classification beyond M to L and T and
 spectroscopic classification criteria relevant to these classes are described. 
 Contemporary methods of classifications based upon different automated
 approaches  are introduced.
\end{abstract}
{\bf Keywords:} Spectral classification, luminosity classes, metal-poor stars,
M-L-T sectral classes
\section{Historical Account of Spectral Classification }
 
\label{sec:1}
\label{1.Historical Account }
 In 1866, Fr Angelo Secchi a Jesuit astronomer working in Italy observed
 prismatic spectra of about 4000 stars visually and divided stars in
 four broad spectral classes using common absorption features of hydrogen.
 During 1886-97, Henry Draper Memorial Survey at Harvard carried out a
 systematic photographic spectroscopy  of stars
 brighter than 9th magnitude covering
  entire sky  using telescopes at Harvard and Arequipa, Peru under the
 leadership of E.C.Pickering. The Henry Draper Catalog  was published
 in 9 volumes of the Annals of Harvard College Observatory between
 1918 and 1924. It contains rough positions, magnitudes,
 spectral classifications for 225,300 stars. Earlier work by W. Fleming
 essentially  subdivided the   previously used Secchi classes 
(I to IV) into more specific classes,  by giving letters from A to N. 
 The strength of hydrogen lines being the main classifier,
 the spectral type A was assigned to stars with strongest hydrogen
 lines followed by B,C with weaker hydrogen lines. This system was 
 found to be unsatisfactory since the line strengths of other lines
 varied irregularly and so did the B-V color. This system was improvised
 by A.Maury, A.J.Cannon and E.Pickering who re-arranged the spectral 
 sequence taking into consideration the changes in other lines
 and this new spectral sequence was also a sequence according to the
 color of the stars. But well-known stars had been already  assigned the older
 spectral classes for long hence it was not possible to change them.
 We, therefore, have a spectral sequence essentially temperature
 dependent but goes like OBAFGKM.
                                  At the cool end the classification 
 becomes more complex with parallel branches of R,N,S stars. While
 the M stars have TiO bands, S stars display ZrO bands, while R and N 
 are carbon stars showing strong bands of various molecules with 
 carbon. These have more recently been merged into a unified
 carbon classifier C scheme of C1, C2 etc with the old N0 starting at
 roughly C6. Another subset of cool carbon stars are the J-type stars, 
 which are characterized by the strong $^{13}$CN  molecules in addition
 to those of $^{12}$CN.

 Each of the above mentioned spectral classes OBAFGKM have been subdivided
 into ten subclasses e.g A0, A1 ... A9.

\subsection{Luminosity Effects in Stellar Spectra}

 E. Hertzsprung suggested in 1905 that spectral line widths were related
to the luminosity of the stars. He pointed out that at a given apparent 
magnitude, the low proper motion stars would  be at larger
distance from us than the high proper motion stars of the same apparent 
magnitude and hence of higher intrinsic luminosity. These low proper
motion stars were found to be exhibit narrower spectral lines so
Hertzsprung concluded that these narrow line stars have larger
intrinsic luminosity than the broad line stars.

 In 1943, W.Morgan, P.C. Keenan and E.Kellman introduced luminosity
 as second classification parameter. Morgan noticed the near constancy of 
 the gravity along the main sequence in HR diagram and luminosity class
 parameter was an attempt to identify stars of different gravities and
hence radii at nearly constant temperature.

 The above mentioned system also known as Yerkes Spectral Classification. 
 Within the system, six luminosity classes are defined on the basis
 of standard stars over the  observed luminosity range. 

The Six classes are:

Ia: most luminous supergiants 

Ib: less luminous supergiants

II: luminous giants

III: normal giants

IV: subgiants

V: main sequence stars

The main  sequence class (dwarfs) are the stars at the main sequence,
 sustaining themselves through the conversion of hydrogen to helium by nuclear fusion
 in the stellar core, giant is a post main sequence star which is no 
longer burning hydrogen at the core but has H-burning shell outside
the core. Giants as well as supergiant have comparable mass to dwarfs
but have expanded to a much larger radii resulting in a decrease in their
surface gravities. The spectral lines appear  broad in the dense
atmospheres of dwarfs primarily due to pressure broadening and Stark 
broadening while the same line would appear narrow in the low
gravity atmospheres of supergiants.

 The luminosity effects
are not restricted to the narrowing of strong lines. The line strengths
and ratios of line strengths of neutral and ionized species also show
remarkable variations over spectral classes and luminosity types and
have been used for defining the subclasses and luminosity types. In addition,
there are well-known luminosity indicators such as the
emission components in the lines of CaII 
H and K  in late type stars which are
related to the luminosity (absolute magnitude) of the stars and the
calibration of this relationship has been carried out by O. Wilison and
M.K.V.Bappu in 1957. The strength of near IR OI triplet at 7771-75$\AA$
has been used by Osmer (1972), Arellano Ferro, Giridhar and Goswami (1991)
Arellano Ferro, Giridhar and Rojo Arellano  (2003)
for absolute magnitude calibration of A-G stars.  In the next section, we will
describe the line strengths and their ratios which are used to define
 the spectral classes and luminosity types.

\begin{figure}
\centering
\includegraphics[height=8cm]{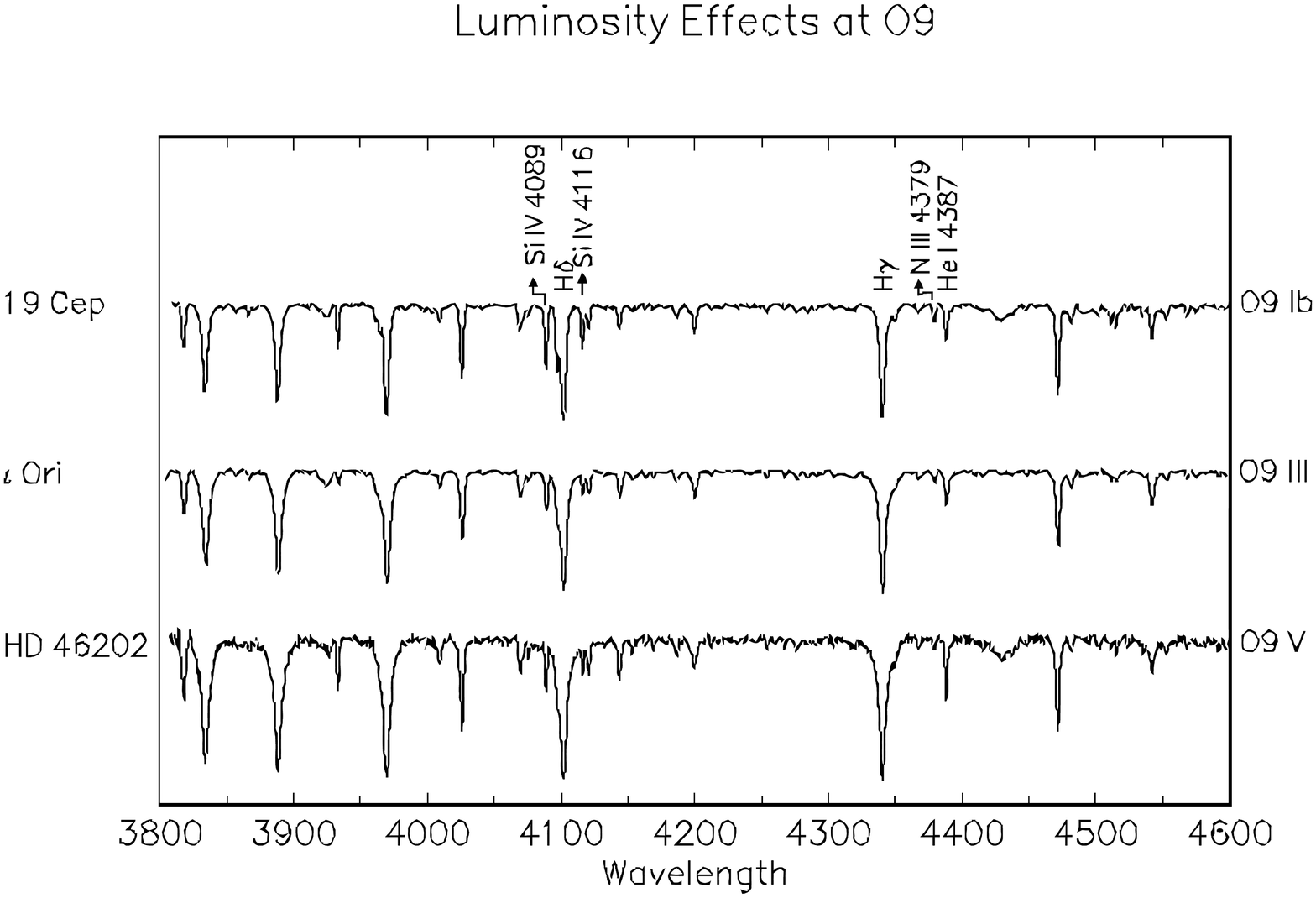}
\caption{ The luminosity effect in late O stars from Gray 2003 with author's permission}
\label{fig:1}       
\end{figure}
\clearpage

\section {Classification Criteria for various spectral types }
\label{sec:2}
\label{2.Classification Criteria for various spectral types }
Beginning from O stars which are hottest, we will briefly describe the spectral
characteristic for each spectral class and also list the spectral type and
luminosity class indicators. Most of the information for this tutorial section
 is taken from Jascheck and
Jascheck (1987). The figures used in this section illustrating the different spectral types and also the luminosity
effects at various spectral types are taken from "A Digital Spectral Classification
Atlas v1.02 " by R.O.Gray with his permission. More figures can be found on the website
http://nedwww.ipac.caltech.edu/level5/Gray/frames.html.

\subsubsection{ O-type Stars }
These are hot stars with temperature range of 28,000K to 50,000K.
These stars exhibit strong lines of neutral and ionized helium.
 The strength of HeII lines
decrease and HeI as well HI lines increasei in strength towards later O-types.
Over the spectral range O4 to B0 the line strength  of HeII line at 4541\AA~ decreases
 from 800m\AA~ to 200m\AA~, while HeI line at 4471\AA~ increases from 100m\AA~ to
1000m\AA~ and that of HI line at 4341\AA~ increases from 1.5\AA~ to 2.5\AA~.  
The O-type spectra also exhibit the features of doubly and triply ionized
carbon, nitrogen, oxygen and silicon. The line ratios such as CIII$\lambda$ 4649/ 
HeII$\lambda$ 4686 are used for luminosity classification. Similarly, the ratio 
of Si IV $\lambda$ / HeI$\lambda$4143 serves as good luminosity
indicator in late O type stars. 
Wolf-Royet stars are a special family of O-type stars that are characterized
by broad emission lines of ionized helium, carbon (WC sequence) or nitrogen (WN sequence).
The WC stars exhibit emission lines of HeII such as HeII$\lambda$ 4686, 
ionised carbon such as CII$\lambda$4267,
CIII$\lambda$ 3609,4187,4325,4650 etc CIV$\lambda$ 4441,4658,4758 etc and lines of OII,OIII,
OIV and OV. These star are subdivided into WC classes from WC2-WC10 based upon the line
ratios of CIII$\lambda$5696/OV $\lambda$5592, CIII$\lambda$5696 / CIV$\lambda$5805 etc.

The WN stars exhibit emission lines of HeII , NIII$\lambda$4097, 4640,5314, NIV$\lambda$
3483,4057 and NV $\lambda$4605,4622. These stars are also subdivided into WN subclasses 
using the line ratios such as NIII$\lambda$4640 / HeII$\lambda$4686.

\begin{figure}
\centering
\includegraphics[height=8cm]{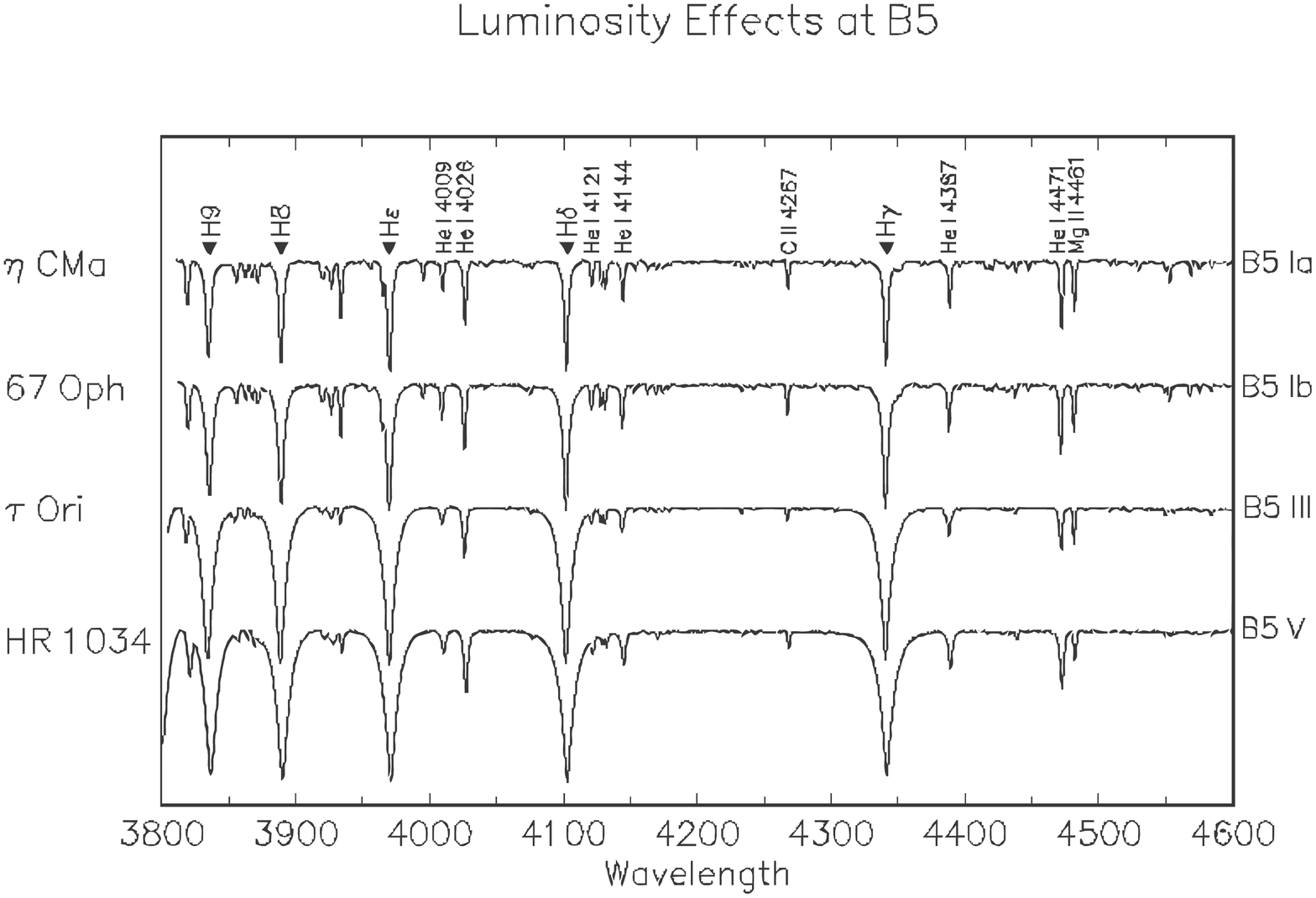}
\caption{ The luminosity effect in B stars is  illustrated above. The
 figure is  from Gray 2003 with author's permission }
\label{fig:2}       
\end{figure}

\subsubsection{ B-type Stars }
The B type spectra contains lines of HeI, HI, CII, CIII, NII, NIII, OII, SiII, SiIII, SiIV, MgII.
The lines of higher ionized states of C,N,O are present in early B stars. The maximum
strength of HeI line reaches near B2.  
Many B stars are fast rotators and a emission lines are present in some of them.

 The line ratios SiIII / SiIV , SiII$\lambda$ 4128-30 / HeI$\lambda$ 4121
and SiII$\lambda$ 4128-30 / HeI$\lambda$ 4144 are used for spectral class determination.
The luminosity criteria used include the ratios of features at $\lambda$4119 (SiIV+HeII)
/ $\lambda$4144 (HeI), $\lambda$4481 (MgII) / $\lambda$4471 (HeI) which increase
with luminosity. Profiles of Balmer lines become narrower with luminosity. 
These are blue white stars with temperature range of 10,000K to 28,000K.

\begin{figure}
\centering
\includegraphics[height=8cm]{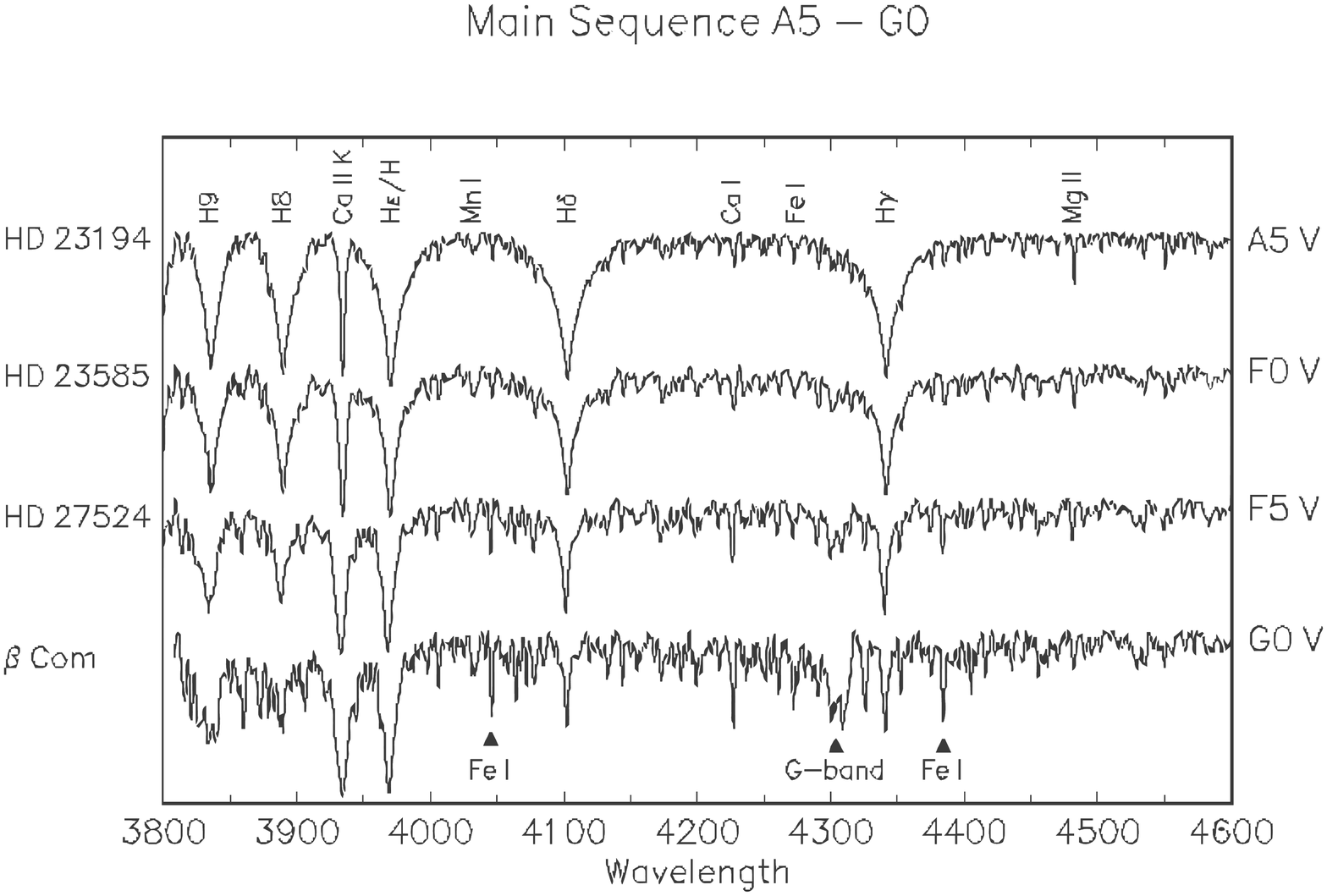}
\caption{ The luminosity effect in A5 to  G0 stars are illustrated the 
 figure is taken from Gray 2003 with author's permission }
\label{fig:3}       
\end{figure}

\subsubsection{ A-type Stars }
These are white stars with temperature range of 7,500K to 10,000K.
The A type stars exhibit strong hydrogen lines of Balmer series. The hydrogen
lines are strongest at spectral type A2. The H$_\gamma$ has a strength
of 13.6$\AA~$ at A0, 17$\AA~$ at A2 and decreases to 13 $\AA~$ at A7 and to
8$\AA~$ at F0. The similar pattern is followed by other lines of
Balmer series. The metallic lines gradually increase in
strength from A0 to A9. The helium lines are absent. 
To assign subclasses the line ratios CaI$\lambda$4227 / MgII $\lambda$4481,
 FeI $\lambda$ 4045 / FeII $\lambda$4173, MgII $\lambda$4481 / FeI  $\lambda$4485
MgII $\lambda$4481/ FeII  $\lambda$4416
 are useful. However other line ratios are also used.
 The luminosity criteria used are blend ratios such as FeI,FeII $\lambda$4383-85 /
MgII $\lambda$4481, FeII $\lambda$4417 / MgII $\lambda$4481  
 which become stronger towards higher luminosity. The hydrogen lines also
 become narrower towards higher luminosity. The additional luminosity indicators are
 SrII $\lambda$ 4215 / CaI $\lambda$ 4226, FeII $\lambda$4351 / MgII $\lambda$4481
 which increase at higher luminosity.
The near infrared OI triplet at $\lambda$ 7771-75 is a very good indicator
of luminosity for A-F stars.

Additional sub-classification of A type stars such as Am. Ap, is done based on their chemical
peculiarities, magnetic fields and rotation and the presence of emission lines in
their spectra.

\begin{figure}
\centering
\includegraphics[height=8cm]{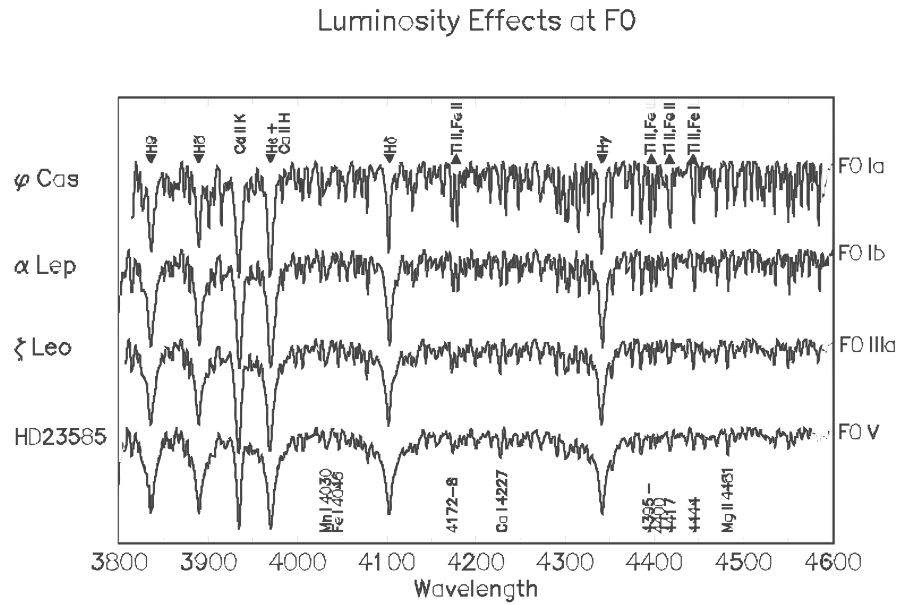}
\caption{ The luminosity effect in F stars is  illustrated, 
 the figure is taken from Gray 2003 with author's permission } 
\label{fig:4}       
\end{figure}

\subsubsection{ F-type Stars }
These are white-yellow stars with temperature range of 6,000K to 7,500K.
The F type spectra have large number of metallic lines, Ca II H and K lines are very
strong which become stronger than hydrogen lines of Balmer series. The hydrogen lines
although present, are on decline from F0- F9. The G band of CH molecule makes its 
appearance around F3. The Ca II at $\lambda$ 3933 increases from 6.5$\AA~$ at F0 to
 17.0$\AA~$ at G0, CaI at $\lambda$ 4226 increases from 0.25$\AA~$ at F0 to
 1.1$\AA~$ at G0 while H$_{\gamma}$ decreases from 8 $\AA~$ at F0 to 3$\AA~$ at G0.
 Due to large number of lines present, at the resolution used for classification,
 most feature are blended. For the spectral class the criteria used are FeI$\lambda$ 4045 /
 H$_{\delta}$, CaI $\lambda$ 4226 / H$_{\gamma}$, MnI $\lambda$ 4030-34 / Si II $\lambda$ 
 4128-32 etc.

The CaII lines show positive luminosity effect,
 ratios Ti II $\lambda$ 4444 /  MgII $\lambda$4481,
 SrII $\lambda$ 4077 / FeI$\lambda$4045 , SrII$\lambda$4077 /  H$_{\delta}$  are
used for luminosity classification.

\subsubsection{ G-type Stars }
These are yellow stars with temperature range of 4,900K to 6,000K.
In G type stars the hydrogen lines are further weakened and become comparable to
the strength of metal lines. Metal lines are stronger and more numerous towards
later G, molecular bands of CH and CN become visible.
The spectral types are obtained by taking the ratio of metal lines with
those of hydrogen lines e.g.  using the raios such as FeI $\lambda$ 4384/ H$_{\gamma}$,
 FeI $\lambda$ 4143 /  H$_{\delta}$,  FeI $\lambda$ 4045 /  H$_{\delta}$, 
 Ca I $\lambda$ 4226 / H$_{\delta}$ etc. For Spectral types later than G5 the
 CaI $\lambda$ 4226 / H$_{\delta}$ can be used. The line ratios such as 
CrI$\lambda$ 4254 / FeI $\lambda$ 4250 or CrI$\lambda$ 4742 / FeI $\lambda$ 4271 
 are recommended for stars showing compositional anomalies such as weak metal line
stars or weak G band stars.

 Luminosity effects at low dispersion can be seen through CN bands. The ratio
 of the SrI+FeI  blend at $\lambda$ 4216 / Ca I $\lambda$ 4226 is known to be
 luminoity sensitive. Ratios of YII+FeI at $\lambda$ 4376 / FeI $\lambda$4383,
 SrII $\lambda$ 4077/  H$_{\delta}$ is also known to be luminosity sensitive
 but will not be suitable for stars with anomalous s-process abundances.
 MgI triplet at 5167-83  $\lambda$ are luminosity sensitive for spectral type
range G8-K5.

\begin{figure}
\centering
\includegraphics[height=8cm]{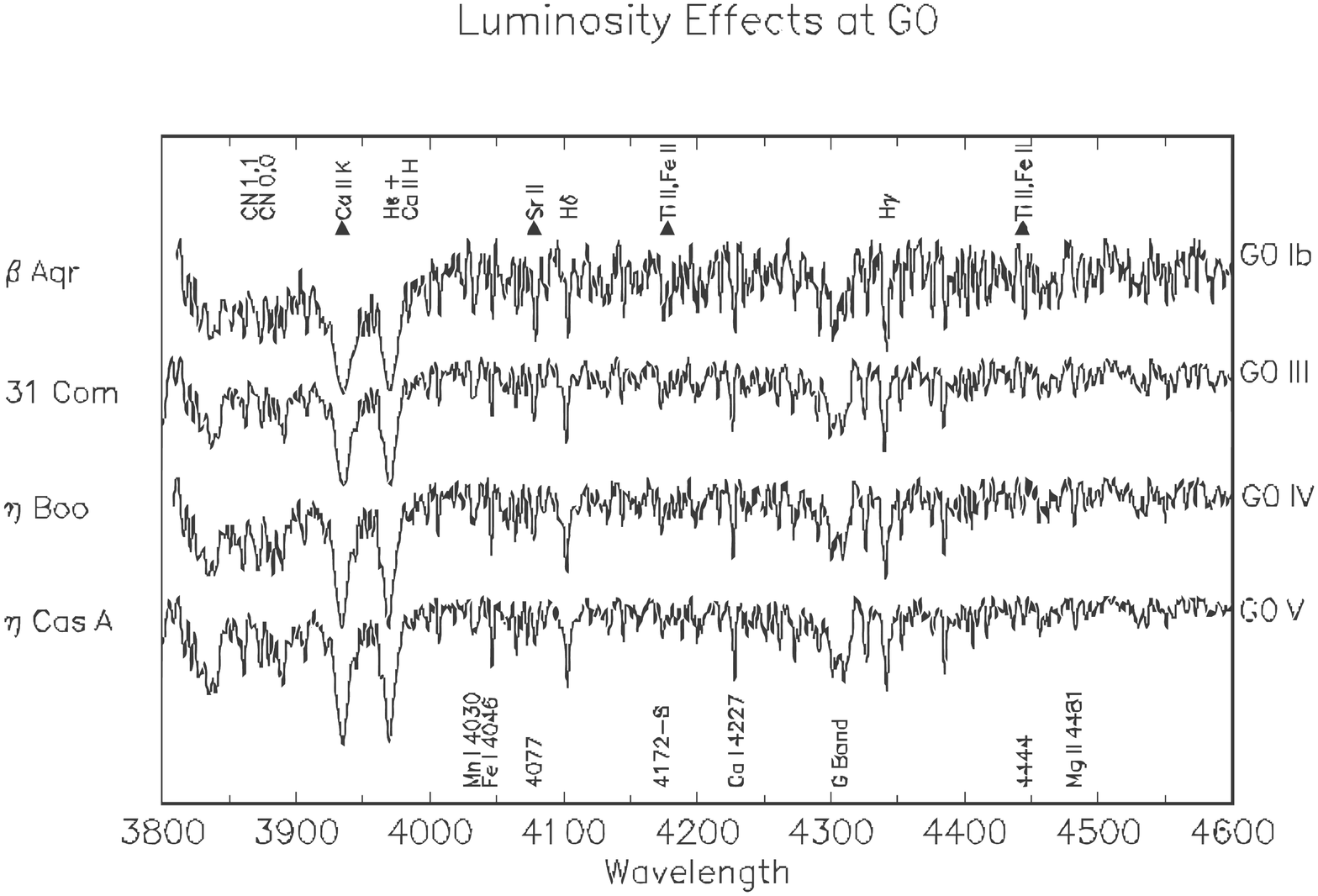}
\caption{ The luminosity effect in G stars is  illustrated 
 the figure from Gray 2003 with author's permission } 
\label{fig:5}       
\end{figure}

\subsubsection{ K-type Stars }
These are orange stars with temperature range of 3,500K to 4,900K.
In these stars the hydrogen lines are very weak but strong numerous  metal
lines are seen. The CaII lines are very strong and CH  molecular band like G band
  becomes very strong. In late K stars TiO and VO bands are also seen. 
 
 The line ratios for spectral type used are 
CrI$\lambda$ 4254 / FeI $\lambda$ 4250, CrI$\lambda$ 4254 / FeI $\lambda$ 4260,
CrI$\lambda$ 4254 / FeI $\lambda$ 4271. Additional qualifiers are     
 TiI$\lambda$ 3999 / FeI $\lambda$ 4005, FeI$\lambda$ 4144 / H$_{\delta}$,
CaI $\lambda$ 4226 / FeI $\lambda$ 4250. The TiO band becomes visible at K7
and MgH at K5. 
 
 The CN band increases with luminosity, so does H$\lambda$ 4101 / FeI $\lambda$ 4071. 
  SrII $\lambda$ 4077 / FeI $\lambda$ 4063, TiII  $\lambda$ /4400,08 /FeI $\lambda$ 4405
 also increase with luminosity.

\subsubsection{ Carbon Stars }

A carbon star is a late type giant with strong bands of carbon compound but no 
metallic oxide band. In their spectra, very intense bands of C$_{2}$, CN and CH 
are present but no bands of TiO, VO are seen.
 The carbon stars as a group have been studied
by Secchi(1868) although in older classification they were given two different
types R and N. The R type stars were similar to late G and early K but exhibited
very strong Swan C$_{2}$ band around  $\lambda$ 4700 and  $\lambda$4395 which is
as strong as G-band of CH. In N stars the the Swan band is so strong that the 
spectrum appears chopped up into the section of different intensities.
Although based upon CN band (at $\lambda$ 4216 and 3833) strengths
 a sequence R0,R1 to R10=N0 and subsequently upto N7 can be defined,
 but the same pattern is not shown by Swan C$_{2}$ bands, which become strongest
 at R5 and weakest at N0 and further strengthening towards later N types.
 A more detailed study by Shane(1928) 
 revealed that temperature variations over the subclasses are not large
and  branching of these stars R and N stars is caused by abundance difference
in C and O. If oxygen is more abundant than carbon then the spectrum
is dominated by oxides like TiO (M stars).  Keenan and Morgan (1941) used
 the term "carbon stars" 
instead of R and N wherein the overabundance of carbon varied from the star to star.
They established the sequence of carbon stars by means temperature index of
CrI$\lambda$ 4254 / FeI $\lambda$ 4250 and the strength of resonance lines of
NaI at $\lambda$ 5890-96 which are good temperature indicators for the late spectral
type stars. The temperature sequence C0 to C7 covers full range of R0 to  N7
  which in temperature  is very similar to the sequence of G4-M4.
 J stars are another group of carbon stars characterized by unusually strong
 isotopic bands of carbon which implies very low C$^{12}$/C$^{13}$ ratio.

\begin{figure}
\centering
\includegraphics[height=8cm]{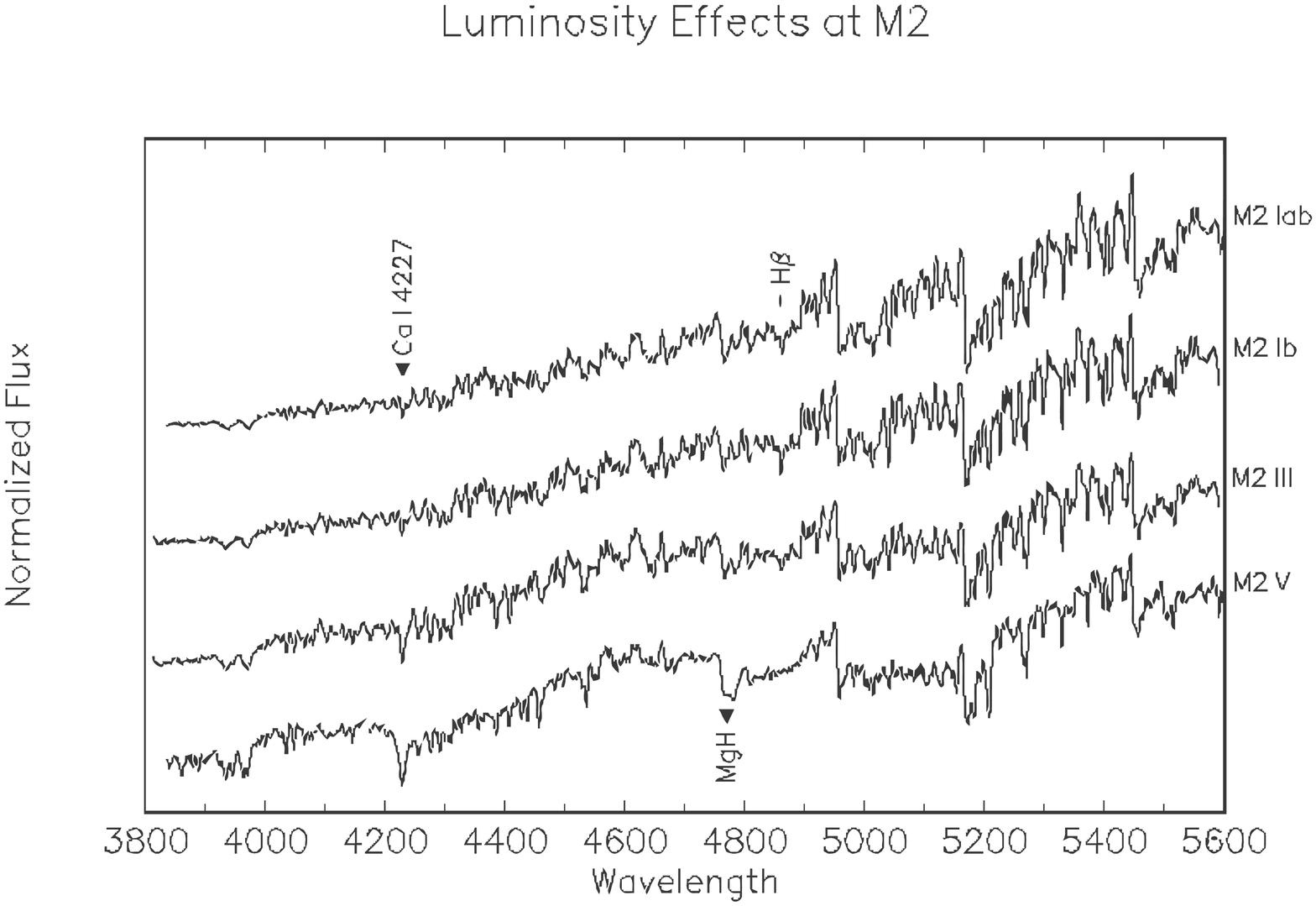}
\caption{ The luminosity effect in M2 stars  is illustrated 
 the figure is  from Gray 2003 with author's permission } 
\label{fig:6}       
\end{figure}
\clearpage

\subsubsection{ M-type Stars }
These are red stars with temperature range of 2,000K to 3,500K.
The spectra of M stars are dominated by strong bands of TiO, VO, LaO. 
The bands used for classification upto M2 are those of
 TiO at$\lambda$ 4584, 4761, 4954 5448. After M3 these
bands saturate, those at $\lambda$ 5759, 5810 saturate after M5. The VO bands
at $\lambda$ 5737, 7373, 7865, 7896 become conspicuous after M7.

The luminosity effect in early M type stars can be seen through 
decrease of CaI $\lambda$ 4226 at higher luminosity; a similar negative
luminosity dependence is exhibited by Cr I feature at $\lambda$ 4254-74-90.
Ratios  SrII $\lambda$ 4077 / FeI $\lambda$ 4263 and (YII +FeI )$\lambda$ 4376 /
FeI $\lambda$ 4386 increase with luminosity. The KI $\lambda$ 7699, NaI $\lambda$ 8183
and $\lambda$ 8185 decrease in intensity from dwarfs to giants. The Ca II triplet
at $\lambda$ 8498, 8542 and 8662 is a very important luminosity indicator
 in these stars. It is very weak in dwarfs but becomes very strong in giants and
 supergiants.
 
\subsubsection{ S-type Stars }
These stars are similar in temperature to K5 to M stars but exhibit strong
bands of ZrO at with band heads at $\lambda$ 4640, 5551 and 6474.

\section{New Spectral types L  and T }
\label{sec:3}
\label{3.New Spectral types L  and T  }

These are the coolest stars with temperatures less than 2500K.
Since they are very cool, they emit mostly in infrared wavelengths. 
The original classification scheme of L dwarfs based on red to far-red spectral
region has been described in Kirkpatrick et al. (1999,2000).
The L stars have a temperature range of 2500K to 1300K and exhibit some
overlapping features with M stars such as TiO, VO although these features do not
remain as strong. These features are very weak in L3 and almost disappear at L6
leaving only metal hydride  bands of FeH, CaH, CrH.  These bands have maximum 
strength at mid L but become weak in late L types where H$_{2}$O  presents
a strong feature at 9300$\AA~$. The atomic features such as resonance lines of RbI
at $\lambda$7800, 7948 $\AA~$ and CsI$\lambda$8521, 8945$\AA~$ can be seen in early L and grow in
strength throughout the sequence. The resonance line of LiI at 6708$\AA~$ is present
in L dwarfs.  The KI resonance doublet shows very remarkable change. Being somewhat
narrower and weak at M9, it broadens and become strong through L0 onwards such that
at L5 two separate cores are indistinguishable and doublet looking like
 a broad trough 
 is  $\sim 600\AA~$ wide; at L8 it is $\sim 1000\AA~$ wide. The NaI resonance doublet
also follows the same pattern by becoming broad and shallow at L5 .

These spectral variations like disappearance of TiO, VO and other molecular features
at later L subclasses are believed to be caused by formation of various types of
condensates at the temperatures lower than 2600K. Lodders (1999) have made extensive
chemical equilibrium calculations and the atmospheric composition changes as the
material is removed from gas to solid phase. 

\begin{figure}
\centering
\includegraphics[height=10cm]{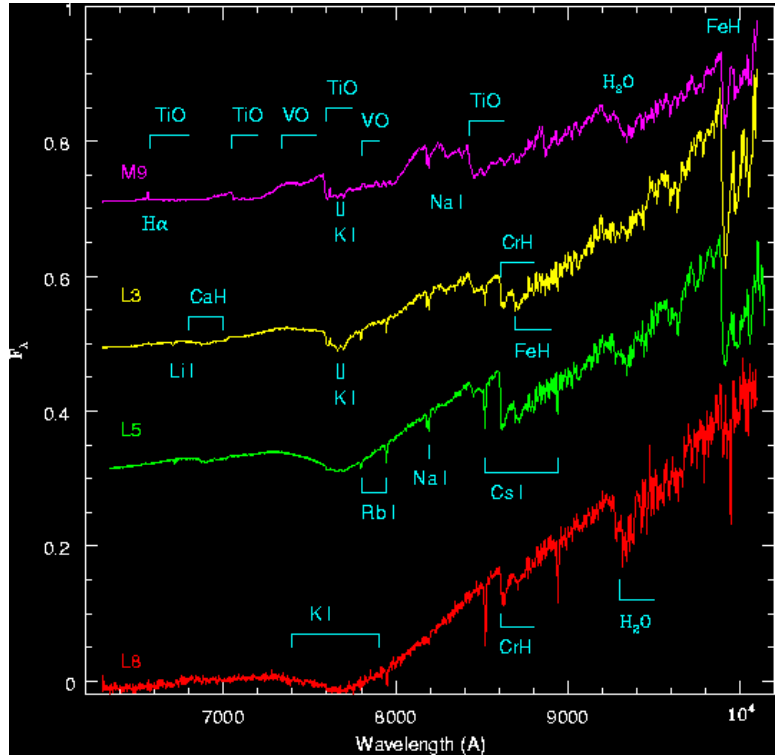}
\caption{ The  spectral variations seen from late M to L stars are illustrated.
The figure is taken from the website www.stsci.edu/inr/ldwarfs originally published
in Kirkpatrick 1999, ApJ {\bf 519}, 802, by permission of  the AAS}
\label{fig:7}       
\end{figure}

\subsection{The T dwarfs}
As the brown dwarfs cool to temperatures below 1400K, a drastic change occurs
between 1400K to 1200K  when CH$_{4}$ bands become strong while CO band weaken.
The CH$_{4}$ bands are extremely broad  in near infrared and they
 even modify the near infrared broad band colors. Strong methane absorption at 
 1.6 and 2.2 $\mu$  reduces more than half flux from H and K pass-band giving them
 unrealistic colors.

\begin{figure}
\centering
\includegraphics[height=10cm]{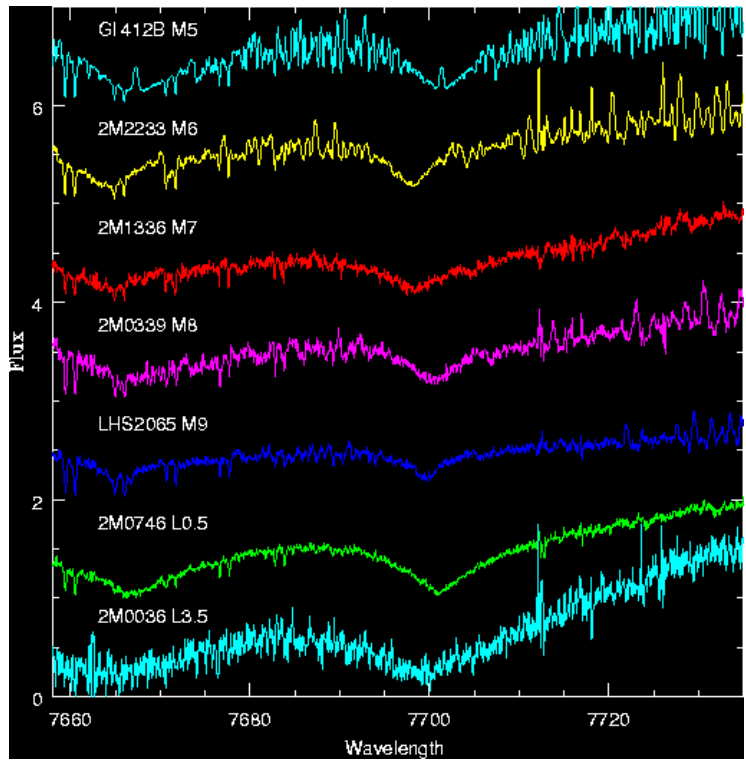}
\caption{ The drastic variations in KI resonance doublet is illustrated above.
The figure is  originally published in
 Reid, Kirkpatrick 1999, AJ {\bf 119}, 369, by permission of  the AAS }
\label{fig:8}       
\end{figure}
\clearpage

 A set of flux ratios measuring the strengths of methane and H$_{2}$O have been
 attempted to classify the T dwarfs by Burgasser et al (2002) and others.
 Considerable progress will be made through deeper IR surveys.

\begin{figure}
\centering
\includegraphics[height=10cm]{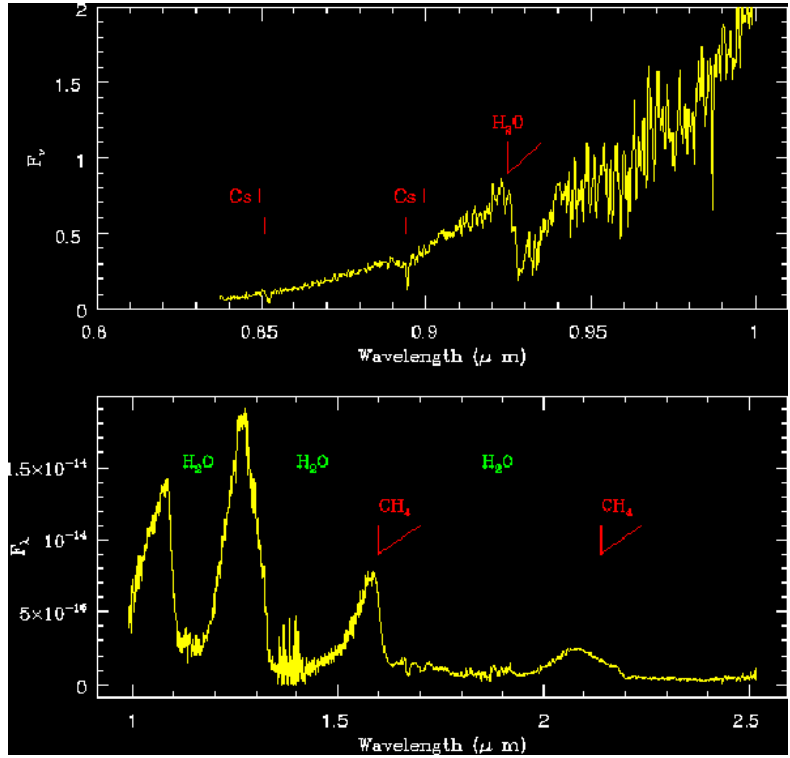}
\caption{ The strengthening of CH4 bands in T stars is  illustrated above.
 The figure was originally published in Oppenheimer et al. 1998, ApJ,{\bf 502},932,
 by permission of  the AAS}
\label{fig:9}       
\end{figure}
\clearpage

\section{Modification of MK system }
\label{sec:4}
\label{4.Modification of MK system }

The MK system has gone through extensive research and refinement. The important work
in refining and extending the MK system in dimensions beyond the traditional two
dimensional  spectral type luminosity class grid has been carried out by Keenan
and co-workers and the use of third "metallicity parameter" is recommended.
Gray (1989) has carried out the extension of MK system to metal-weak F and G stars.
A grid of metal-poor stars standard stars has been prepared with the objective of
providing a mean of classifying metal-weak stars using a standard of similar 
metallicity rather than the old method of comparing the metallic line strengths of
the program stars with that of a considerable earlier solar composition standard.

 The earlier attempt of deriving spectral types for metal-poor stars relied upon
 hydrogen line strengths or G-band strength due to their strong temperature 
 dependence. However these classifiers are not adequate; the hydrogen line strengths
 do depend upon other stellar parameters and are not suitable at the later
 spectral types.   Corbally (1987) pointed out that the ratio of CrI resonance 
 triplet at 4254, 4274, 4290 $\AA~$ to the neighboring lines of FeI at 4251, 4272,
4362 $\AA~$ arising from excited level is nearly independent of metallicity and
 hence can serve as temperature indicator. 
 For luminosity classification the ratios of features like  Ti II, Fe II $\lambda$ 4172-9 /
 Fe I $\lambda$ 4203, $\lambda$ 4271 is useful for  F stars while for G stars 
 SrII $\lambda$ 4077/ FeI $\lambda$ 4063, 4046 and Fe,YII $\lambda$ 4373/ 4383 features
 may be useful since these lines (blends) are very strong and  remain
 discernible even at low metallicity. A list of standard stars for metallicity classification
 can be found in Gray(1989).
\begin{figure}
\centering
\includegraphics[height=8cm]{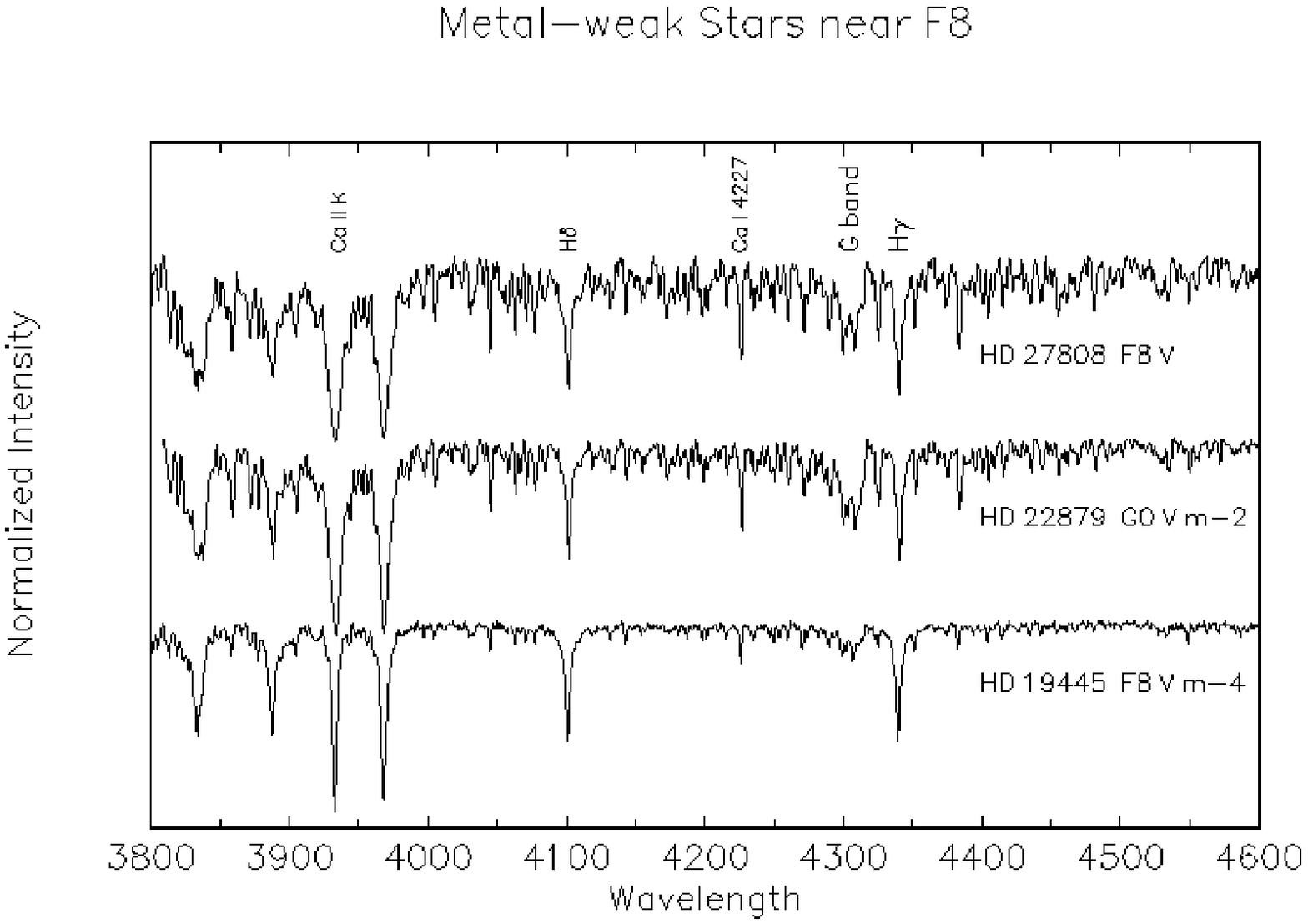}
\caption{ The spectra of a few metal-poor stars is shown above}
\label{fig:10}       
\end{figure}
\clearpage
 
\section{Contemporary methods of spectral classification }  
\label{sec:5}
\label{5.Contemporary methods of spectral classification }  
Stellar classification has been an important tool in stellar and Galactic astronomy
since it provides empirical measure of the fundamental stellar parameters such
as the temperatures, luminosity and metallicity.

 The massive surveys both ground based as well from space missions provide 
 large number of stellar spectra covering  distant components of Galaxy.
 To understand the complex evolutionary history of our Galaxy,
 rapid and accurate methods
 of stellar  classification are necessary.
A short review of the automated procedures are presented here.
 The most commonly used automated spectral classification methods are
 based on (a) Minimum Distance Method (MDM) (b) Gaussian Probability Method
 (GPM), (c) Principal Component Analysis (PCA) and (d) Artificial Neural
 Network (ANN).
 It is beyond the scope of this talk to explain all of them. We chose to
 describe only two of them to introduce the automated approach of classification. 

 In MDM, the classification is done by minimizing the distance metric between the
 object to be classified and each member of a set of templates. The object
 is assigned the class of the template, which gives the smallest distance.
 In this approach the number of
 templates used to define subclasses limit the accuracy of classification.
 Interpolation can be made to make intra-class assignment.
  Katz et al. (1998) used this method with $\chi$$^{2}$ minimizing
 on high resolution
  Elodie spectra  using a large number of reference stars of known T$_{eff}$,
  log $g$ and [M/H] to derive atmospheric parameters of target stars. These
  authors achieved accuracy of 86 K in T$_{eff}$, 0.28 in  log $g$ and 0.35 in
  [M/H]. 

 The Neural Network methods have become very popular due to their speed
 and objectivity. As explained in the papers by Bailer-Jones et al. (1998,2002),
 Ted von Hippel et al. (1994),
 Singh et al.(2002) and more recently in Giridhar, Muneer \& Goswami (2006),
 is a computational method which can provide non-linear
 parametrized mapping between an input vector (a spectrum for example)
 and one or more outputs like SpT, LC or  T$_{eff}$, log $g$ and [M/H].
 For the network to give required
 input-output mapping, it must be trained with the help of representative
 data patterns. These  could be stellar spectra ( or a set of line strengths
 measured from a spectrum)  for which classification or
 stellar parameters are well determined. The training procedure
  is a numerical least
 square error minimization method. The training proceeds by optimizing the network
 parameters (weights) to give minimum classification error. Once the network
 is trained the weights are fixed, the network can be used to produce
 output  SpT, LC or  T$_{eff}$, log $g$ and [M/H] for an unclassified  spectrum.

 The ANN has been used in very large number of stellar applications.
  Vieira and Ponz (1995) have used ANN on low-resolution IUE spectra
  and have determined SpT with an accuracy of 1.1 subclass. 
 Bailer-Jones, Irwin and von Hippel (1998) used ANN to
  classify spectra from Michigan Spectral Survey with an accuracy
  of 1.09 SpT. Allende Prieto et al. (2000) used ANN in their
 search of metal-poor stars.  Snider et al. (2001) used ANN for
 the three dimensional classification  of metal-poor stars.

 We have made a modest effort to use ANN for parametrization of a sample
    of stars in temperature range 4500 to 8000 K. We have used a
    medium resolution Cassegrain spectrograph with the  2.3 m Vainu Bappu
    Telescope at VBO, Kavalur, India to get spectra at resolution (R) of 2000.
  Using the 90 spectra for stars of known parameters, and ANN of
680:11:3 architecture we could attain an accuracy of $\pm$200K in
 temperature and $\pm$0.3dex in metallicity.

  It is very important to envisage an approach that would
   give quick, reliable spectral classifications (or stellar
   parameters) for stars falling in all regions of HR diagram.
 The pipeline procedures are being
developed for the future ambitious missions such as Gaia, Pan-Stars.
    
%
%


\printindex
\end{document}